\documentclass[aps,prd,showpacs,preprintnumbers,amssymb,floatfix]{revtex4}
\usepackage{amsmath}
\usepackage{graphicx}
\usepackage{bm}
\usepackage{color,colordvi}
\usepackage{url}
\begin{document}

\newcommand{\as}{\alpha_{\textrm s}}

\def\plus{{\!+\!}}
\def\minus{{\!-\!}}
\def\z#1{{\zeta_{#1}}}
\def\zz#1{{\zeta_{#1}^2}}

\def\ca{{C^{}_A}}
\def\cf{{C^{}_F}}
\def\tr{{T^{}_{\! R}}}

\def\caca{{C^{2}_A}}
\def\cfcf{{C^{2}_F}}

\def\nf{{n^{}_{\! f}}}
\def\n2f{{n^{\,2}_{\! f}}}

\def\nc{{N^{}_{\! c}}}
\def\n2c{{N^{\,2}_{\! c}}}

\def\li#1{{\textrm{Li}_{#1}}}
\def\h{{\textrm{H}}}
\def\D#1{{\mathcal{D}_{#1}(z)}}

\preprint{FSU-HEP-2007-0501}
\preprint{BNL-HET-07/7}
\preprint{arXiv:0705.0035 [hep-ph]}
\title{Higgs boson production with one bottom quark including
       higher-order soft-gluon corrections}
\author{B. Field}
\email[]{bfield@hep.fsu.edu}
\affiliation{Department of Physics,
             Florida State University, Tallahassee, Florida
             32306-4350, USA}
\author{C.B. Jackson}
\email[]{jackson@quark.phy.bnl.gov}
\affiliation{Physics Department,
             Brookhaven National Laboratory,
             Upton, New York, 11973-5000, USA}
\author{L. Reina}
\email[]{reina@hep.fsu.edu}
\affiliation{Department of Physics,
             Florida State University, Tallahassee, Florida
             32306-4350, USA}
\date{May 1, 2007}

\begin{abstract}

A Higgs boson produced in association with one or more bottom quarks
is of great theoretical and experimental interest to the high-energy
community. A precise prediction of its total and differential
cross-section can have a great impact on the discovery of a Higgs
boson with large bottom-quark Yukawa coupling, like the scalar ($h^0$
and $H^0$) and pseudoscalar ($A^0$) Higgs bosons of the Minimal
Supersymmetric Standard Model (MSSM) in the region of large
$\tan\beta$. In this paper we apply the threshold resummation
formalism to determine both differential and total cross-sections for
$b g \rightarrow b\Phi$ (where $\Phi\!=\!h^0,H^0$), including up to
next-to-next-to-next-to-leading order (NNNLO) soft plus
virtual QCD corrections at next-to-leading logarithmic (NLL)
accuracy. We present results for both the Fermilab Tevatron and the
CERN Large Hadron Collider (LHC).

\end{abstract}

\maketitle

\section{Introduction} 
\label{sec:intro}

The prospect of discovering the mechanism of Electroweak Symmetry
Breaking (EWSB) in the coming years is of primary interest to
high-energy physics programs around the world. In the Standard Model
(SM), a single $SU(2)_L$ complex scalar doublet gives rise to one
Higgs boson ($h$) and massive gauge bosons through the Higgs mechanism
as well as massive fermions through Yukawa interactions. In the
Minimal Supersymmetric Standard Model (MSSM), there are two complex
scalar doublets each giving rise to the masses of the up- and
down-type quarks respectively. After EWSB in the MSSM, five physical
Higgs bosons remain: a light and heavy scalar ($h^0$, $H^0$), a
pseudoscalar ($A^0$), and two charged scalars ($H^\pm$). Their
couplings to the SM particles may substantially differ from the SM
Higgs boson couplings. Many of the properties of the SM and MSSM Higgs
bosons have been reviewed in Refs.~\cite{Gunion:1989we,Carena:2002es}.

There have been some rapid changes in the precision electroweak fits
that have lead to new bounds on the Higgs boson mass in the SM, mostly
from a shift in the central value of the top quark
mass~\cite{Lepewwg:2007}. Together with the lower bound set by direct
searches for a SM Higgs bosons at LEP-2~\cite{Barate:2003sz}, precision
electroweak fits indicate that $114.4 < M_{h} < 166-199$~GeV at the
$95$\% confidence level~\cite{Lepewwg:2006}. The experimental bounds
on the MSSM Higgs bosons are weaker than those of the SM due to the
much larger parameter space of the supersymmetric model, leading to a
$M_{h^0,A^0} > 93$~GeV lower
bound~\cite{Lephwg:2004,Lephwg:2005}. However, the lightest MSSM
scalar is theoretically bounded to lie below about
$130$~GeV. Therefore at least one Higgs boson that may exist in nature
(either SM or MSSM) will probably be constrained by the Tevatron and
definitely probed by the Large Hadron Collider (LHC).

In the SM, for current and future planned energies at hadron
colliders (both $p\bar{p}$ and $pp$), a Higgs boson would be primarily
produced via gluon fusion ($gg \rightarrow h$) through a top-quark
loop. However, in the MSSM the production channels can become
quite varied depending on the parameter space. In particular, for
large values of $\tan\beta$ (the ratio of the two Higgs doublets vacuum
expectation values) the MSSM bottom-quark Yukawa couplings
are enhanced and the production of a Higgs boson with bottom quarks
become the leading production mode. Indeed, this has already been used
at the Tevatron to substantially constrain the MSSM parameter
space~\cite{Abazov:2005yr}.

Bottom quarks produced in association with a Higgs boson have also
raised a great deal of theoretical interest~\cite{Campbell:2002zm,
Maltoni:2003pn,Dittmaier:2003ej,Dawson:2003kb,Campbell:2004pu,
Kramer:2004ie,Dawson:2004sh,Maltoni:2005wd,Dawson:2005vi}. Although
there is a conceptual difference in whether one considers bottom
quarks in the initial state as partons or only gluons and light quarks
(i.e., whether one employs the five flavor number scheme, 5FNS, or the
four flavor number scheme, 4FNS), physical observables have been shown
to agree remarkably well in both schemes once full next-to-leading
order (NLO) QCD corrections are
included~\cite{Campbell:2004pu,Kramer:2004ie,Dawson:2004sh}. This has
placed the theoretical predictions of both total and differential
cross-sections under better control. Further improvement can be
achieved by considering the impact of resumming corrections that can
be relevant in specific regions of phase space.

In this paper we investigate the impact of higher-order QCD threshold
corrections on the $bg\rightarrow b\Phi$ ($\Phi\!=\!h^0,H^0$) total
and differential cross-sections, both at the Tevatron and at the
LHC. The resummation of soft plus virtual dynamical threshold
corrections to a variety of processes involving both scalars and
pseudoscalars Higgs bosons has been studied extensively in the
literature~\cite{Catani:1989ne,Kauffman:1991jt,Yuan:1991we,Kauffman:1991cx,
Catani:1996yz,Kramer:1996iq,Balazs:2000wv,deFlorian:2000pr,
deFlorian:2001zd,Glosser:2002gm,Berger:2002ut,Berger:2003pd,
Bozzi:2003jy,Catani:2003zt,Kulesza:2003wn,Field:2004tt}. In
particular, following the resummation techniques originally proposed
in Refs.~\cite{Collins:1981uk,Collins:1981va,Collins:1984kg}, we will use
the formalism recently outlined in
Refs.~\cite{Kidonakis:2003tx,Kidonakis:2004ib,Kidonakis:2005kz}, to
which we refer for further references.

In order to assess the validity of the threshold resummation formalism
for the case of $bg\rightarrow b\Phi$ ($\Phi\!=\!h^0,H^0$) production,
we compare the NLO full calculation (in the
5FNS)~\cite{Campbell:2002zm} as obtained from MCFM~\cite{mcfm} with
the results obtained by expanding the resummed cross-section to
include NLO soft plus virtual corrections up to next-to-leading
logarithmic (NLL) accuracy.  The NLO-NLL formalism reproduces very
closely the full NLO calculation in the region of small Higgs boson
transverse momentum, where most of the statistics are accumulated. We
then improve upon the NLO fixed-order calculation by including both
next-to-next-to-leading (NNLO) and next-to-next-to-next-to-leading
order (NNNLO) soft plus virtual corrections up to NLL
accuracy. Higher-order corrections are more and more stable with
respect to variations of both the renormalization ($\mu_R$) and
factorization ($\mu_F$) scales, in particular in the threshold region
where soft corrections dominate the bulk of radiative corrections. We
will discuss more extensively the perturbative behavior of NNLO-NLL
and NNNLO-NLL corrections in Section~\ref{sec:results}.

Overall, the NNLO-NLL and NNNLO-NLL effects are sizable and they
greatly stabilize the cross-section. The improved small transverse
momentum behavior should help to shed some light on the current bounds
on $\tan\beta$ in a more robust way, since it corrects the
distribution in a region where most of the statistics are accumulated.

\section{MSSM Parameters}
\label{sec:mssm}

Higgs boson physics when associated with bottom quarks is dominated by
two factors -- the running of the bottom-quark mass and the MSSM
couplings to the different CP-even and CP-odd neutral Higgs
bosons. The one-loop supersymmetric corrections to the bottom-quark
Yukawa coupling must also be considered. We will briefly consider each
of these factors to establish our notation.

The SM bottom-quark coupling to the Higgs boson is
$g^{\textrm{SM}}_{b\bar{b}h} = \overline{m}_b(\mu_R)/v$, where the
vacuum expectation value ($v$) of the Higgs doublet is defined as
$v=(\sqrt{2} G_F)^{-1/2} = 246~$GeV and $\overline{m}_b(\mu_R)$ is the
$\overline{\textrm{MS}}$ running mass of the bottom quark as a
function of the renormalization scale $\mu_R$. We define
$\overline{m}_b(\mu_R)$ up to 4 loops (as we will need for the NNNLO
corrections) according to Ref.~\cite{Vermaseren:1997fq}, which
corresponds at 1 and 2 loops to the following expressions:
\begin{align}
\label{eq:mb_ms_1l_2l}
\overline{m}_b(\mu_R)_{1l} &= m^{\text{pole}}_b
\biggl[ \frac{\alpha_s(\mu_R)}{\alpha_s(m^{\text{pole}}_b)}
\biggr]^{c_0 / b_0}\,\,\,, \\
\overline{m}_b(\mu_R)_{2l} &= m^{\text{pole}}_b
\left[ \frac{\alpha_s(\mu_R)}{\alpha_s(m^{\text{pole}}_b)}
\right]^{c_0 / b_0}
\left[ 1+\frac{c_0}{b_0} \biggl( c_1 - b_1 \biggr)
         \frac{\alpha_s(\mu_R)-\alpha_s(m^{\text{pole}}_b)}{\pi}\right]\,\,\,,
\nonumber
\end{align}
where $\alpha_s(\mu_R)$ is the strong coupling constant at the
renormalization scale $\mu_R$ and,
\begin{align}
\label{eq:b0_b1_c0_c1}
  b_0 &=\frac{1}{4\pi}
        \biggl( \frac{11}{3}\nc - \frac{2}{3} \nf \biggr), \qquad
  c_0  =\frac{1}{\pi}, \qquad
  b_1  =\frac{1}{2\pi}
        \biggl( \frac{51 \nc - 19 \nf}{11 \nc - 2 \nf} \biggr), \qquad
  c_1  =\frac{1}{72\pi}
        \biggl( 101 \nc - 10 \nf \biggr)\,\,\,,
\end{align}
are the first two coefficients of the QCD beta function and mass
anomalous dimension function, with $\nc=3$, the number of colors, and
$\nf=5$, the number of light flavor.  The quantity
$m_b^{\textrm{pole}}$ is the bottom-quark pole mass which we take to
be $m_b^{\textrm{pole}}=4.62$~GeV. The 3-loop and 4-loop expressions
of $m_b(\mu_R)$ are lengthy and we refer to
Ref.~\cite{Vermaseren:1997fq} for their exact definition. In the
following, we rescale the 5FNS NLO results obtained from existing
calculations via MCFM in such a way to agree with this definition
of $\overline{m}_b(\mu_R)$. This is possible because
$\overline{m}_b(\mu_R)$ enters only in the overall bottom-quark Yukawa
coupling.

The tree level CP-even neutral Higgs boson couplings to bottom quarks
in the MSSM can be written in terms of the SM couplings as,
\begin{equation}
\label{eq:b_yukawa_mssm_tree}
g^{\textsc{MSSM}}_{b\bar{b}h^{0}} = -\frac{\sin\alpha}{\cos\beta}
g^{\textsc{SM}}_{b\bar{b}h}\,\,\,,
\qquad
g^{\textsc{MSSM}}_{b\bar{b}H^{0}} = \frac{\cos\alpha}{\cos\beta}
g^{\textsc{SM}}_{b\bar{b}h}\,\,\,,
\end{equation}
and are enhanced in the limit of large $\tan\beta$, where $\tan(\beta)
= v_2/v_1$ is the ratio of the vacuum expectation values of the Higgs
doublets coupling to the up- and down-quarks respectively, while
$\alpha$ is the angle which diagonalizes the neutral Higgs sector of
the MSSM. The CP-odd neutral Higgs boson coupling to bottom quarks is
also enhanced exactly by a factor of $\tan\beta$. In the following, we
will focus our discussion on the CP-even neutral Higgs boson cases. In
the $m_b \simeq 0$ limit, the results obtained for the neutral scalars
can be directly rescaled to obtain results for the neutral
pseudoscalar. For non-zero $m_b$, the difference between the scalar
and pseudoscalar case, modulus a rescaling of the couplings, is of
order $(m_b/M_\Phi)^2$ ($\Phi\!=\!h^0,H^0$), and one therefore expects
it to be small.

Finally, we need to consider the supersymmetric corrections to the
bottom-quark Yukawa coupling in Eq.~(\ref{eq:b_yukawa_mssm_tree}) due
to squark and gluino loops. For CP-even neutral Higgs bosons we
find~\cite{Carena:1999py},
\begin{align}
\label{eq:b_yukawa_mssm_1loop}
g^{\textsc{MSSM}}_{b\bar{b}h^{0}} &= 
- g^{\textsc{SM}}_{b\bar{b}h} 
\frac{1}{1 + \Delta_b} 
\biggl[ 
\frac{\sin\alpha}{\cos\beta} - \Delta_b \frac{\cos\alpha}{\sin\beta}
\biggl]\,\,\,, \\
g^{\textsc{MSSM}}_{b\bar{b}H^{0}} &= 
\,\,\,\, g^{\textsc{SM}}_{b\bar{b}h}
\frac{1}{1 + \Delta_b}
\biggl[
\frac{\cos\alpha}{\cos\beta} + \Delta_b \frac{\sin\alpha}{\sin\beta}
\biggl]\,\,\,,
\end{align}
where the one-loop $\Delta_b$ correction can be written as,
\begin{equation}
\label{eq:deltab}
\Delta_b = \mu\tan\beta
\biggl[
\frac{2\alpha_s(m_t)}{3\pi} \,
m_{\tilde{g}} \,
I(m_{\tilde{b}_{1}}, m_{\tilde{b}_{2}}, m_{\tilde{g}}) +
\biggl( \frac{g_{t\bar{t}h}}{4\pi} \biggr)^2
A_t \,
I(m_{\tilde{t}_{1}}, m_{\tilde{t}_{2}}, \mu)
\biggr]\,\,\,,
\end{equation}
and the integral quantity $I(a,b,c)$ is defined as,
\begin{equation}
\label{eq:i_abc}
I(a,b,c) = 
\frac{a^2b^2\ln(a^2/b^2)+b^2c^2\ln(b^2/c^2)+c^2a^2\ln(c^2/a^2)}
     {(a^2-b^2)(b^2-c^2)(a^2-c^2)}\,\,\,.
\end{equation}

Other MSSM parameters of influence in the previous set of equations
are the Higgs-Higgs coupling in the super-potential, $\mu$, the masses
of the up- and down-type squarks after mixing, $m_{\tilde{b}_{1,2}}$
and $m_{\tilde{t}_{1,2}}$, the mass of the gluino, $m_{\tilde{g}}$,
the SM top-quark Yukawa coupling $g_{t\bar{t}h} = m_t/v$, and the
top-quark tri-linear coupling, $A_t$. We set $m_t = 172.2$~GeV,
$m_{\tilde{b}_{1,2}} = m_{\tilde{t}_{1,2}} = M_{\textsc{susy}} =
1$~TeV, $m_{\tilde{g}} = 1$~TeV, $A_b = A_t = 2$~TeV, and $\mu = M_2 =
200$~GeV, where $M_2$ is the $SU(2)$ gaugino mass parameter. With
these values we find $\Delta_b = 0.178$. In our convention, the sign
of Higgs-Higgs coupling, $\mu$, is preferred positive. All of the MSSM
couplings were calculated using the FeynHiggs 2.5
package~\cite{Hahn:2006np} which includes all known corrections to the
bottom-quark couplings in the MSSM through 2 loops.

\section{Resummation}
\label{sec:resummation}

There are several formalisms for resummation calculations. Much of the
research in resummation focuses on total cross-sections, however there
are several ways of approaching differential cross-sections as
well~\cite{Kauffman:1991cx,Berger:2002ut,Kulesza:2003wn,Field:2004tt}.

Different formalisms lend themselves better to different processes. In
general terms, for differential quantities, we can consider the
resummation of one particle inclusive (1PI) or pair invariant mass
(PIM) observables. In the case of our observable, we are interested in
resumming a $2\rightarrow 2$ process, $bg\rightarrow b\Phi$
($\Phi\!=\!h^0,H^0$), so we are in the domain of the 1PI dynamics. The
formalism that we will use to resum threshold corrections to the
differential cross-section of a $2\rightarrow 2$ process, implementing
known universal corrections, is due to
Kidonakis~\cite{Kidonakis:2003tx,Kidonakis:2004ib,Kidonakis:2005kz}.

The calculation of both total and differential cross-sections in
hadron-hadron collisions can be formalized as
\begin{equation}
\label{eq:sigma_had}
\sigma = \sum_{i,j} \int dx_1 \int dx_2
\phi_{i/h_1}(x_1,\mu_F,\mu_R)\phi_{j/h_2}(x_2,\mu_F,\mu_R)
\hat{\sigma}_{ij}(\hat{s},\hat{t}_k,\mu_F,\mu_R)\,\,\,,
\end{equation}
where $\sigma$ and $\hat{\sigma}_{ij}$ could be a total or
differential cross-section of interest, at the hadron and parton level
respectively, while the indices $i$ and $j$ run over the species of
partons contributing to a given process. $\phi_{i/h_l}$ (for
$l=1,2$) is the Parton Distribution Function (PDF) for parton $i$
carrying a fraction $x_l$ of the momentum of hadron $h_l$, at a
factorization scale $\mu_F$ and renormalization scale $\mu_R$. The
partonic center of mass energy is $\hat{s}$ and $\hat{t}_k$ are
partonic $t$-channel type Mandelstam variables.

Higher-order QCD corrections to total and differential parton-level
cross-sections often contain so-called plus distributions and delta
functions of the kinematic variables that in essence measure the
deviation from the kinematic threshold. If we define $\hat{s}_2$ in
terms of the partonic Mandelstam variables as $\hat{s}_2 = \hat{s} +
\hat{t} + \hat{u} -
\sum_i m_i^2$, which vanishes at threshold, we will find plus
distributions $\mathcal{D}_l(\hat{s}_2)$ and delta functions
$\delta(\hat{s}_2)$ in the differential cross-section where,
\begin{equation}
\label{eq:Dl}
\mathcal{D}_l(\hat{s}_2) \equiv
\biggl[ \frac{\ln^l(\hat{s}_2/M^2)}{\hat{s}_2}
\biggr]_+\,\,\,,
\end{equation}
and $M$ denotes a typical mass-scale of the process. At order
$\alpha_s^n$, we expect $l \le 2n-1$ plus distributions. When the
differential or total cross-sections are resummed in moment space,
these plus distributions appear as logarithmic divergences in terms of
the moment variable. These are the logarithms that are resummed at all
orders by the threshold correction resummation formalism. Close to
threshold, i.e. in the region where any extra emitted parton is
necessarily soft, these corrections can be large and even dominate
the cross-section. Hence the need to resum them.

The formalism in Ref.~\cite{Kidonakis:2003tx,Kidonakis:2005kz} allows
us to write these resummed corrections in terms of the tree level
cross-section times universal coefficients based on the color flow of
the process. The color flow of our process, $bg\rightarrow b\Phi$
($\Phi\!=\!h^0,H^0$), is the same as charged Higgs production,
$bg\rightarrow H^-t$, that has been studied in
Refs.~\cite{Kidonakis:2003tx,Kidonakis:2004ib,Kidonakis:2005kz}.

The resummed cross-section cannot be easily evaluated outside of its
moment space~\cite{Kidonakis:2003tx}, however, it can be expanded and
evaluated in the usual manner. Expanding the resummed cross-section to
include NLO, NNLO and NNNLO soft QCD corrections at the NLL accuracy
provides us with the expressions we will use in this paper. According
to Refs.~\cite{Kidonakis:2003tx, Kidonakis:2004ib, Kidonakis:2005kz},
the corrections specific to our process can be written in terms of
three coefficients, $c_1, c_2,$ and $c_3$, determined by the color
flow and the kinematic invariants of the process. For completeness,
let us explicitly give $c_1$, $c_2$ and $c_3$ in the following
equation, using the notation of Ref.~\cite{Kidonakis:2004ib}:
\begin{eqnarray}
\label{eq:c1_c2_c3}
c_1^{bg\rightarrow b\Phi}&=&
\left[C_F\ln\left(\frac{-\hat{u}+M_\Phi^2}{M_\Phi^2}\right)
     +C_A\ln\left(\frac{-\hat{t}+M_\Phi^2}{M_\Phi^2}\right)
     -\frac{3}{4}C_F-\frac{\beta_0}{4}\right]
     \ln\left(\frac{\mu_F^2}{M_\Phi^2}\right)
+\frac{\beta_0}{4}\ln\left(\frac{\mu_R^2}{M_\phi^2}\right)\,\,\,,
\nonumber\\
c_2^{bg\rightarrow b\Phi}&=&T_2^{bg\rightarrow b\Phi}
-(C_F+C_A)\ln\left(\frac{\mu_F^2}{M_\Phi^2}\right)\,\,\,,
\nonumber\\
c_3^{bg\rightarrow b\Phi}&=&2(C_F+C_A)\,\,\,,
\end{eqnarray}
where $T_2^{bg\rightarrow b\Phi}$, the scale-independent part of the
$c_2^{bg\rightarrow b\Phi}$ coefficient, is defined by:
\begin{equation}
\label{eq:t2}
T_2^{bg\rightarrow b\Phi}=
2\mathrm{Re}\Gamma^{\prime(1)}_S-C_F-C_A
-2C_F\ln\left(\frac{-\hat{u}+M_\Phi^2}{M_\Phi^2}\right)
-2C_A\ln\left(\frac{-\hat{t}+M_\Phi^2}{M_\Phi^2}\right)\,\,\,,
\end{equation}
and $\Gamma^{\prime(1)}_S$, related to the one-loop soft anomalous
dimension which describes non-collinear soft gluon emission, is given
by:
\begin{equation}
\label{eq:gamma_prime}
\Gamma^{\prime (1)}_S=
C_F\ln\left(\frac{-\hat{t}+m_b^2}{m_b\sqrt{\hat{s}}}\right)
+\frac{C_A}{2}\ln\left(\frac{-\hat{u}+m_b^2}{-\hat{t}+m_b^2}\right)
+\frac{C_A}{2}(1-i\pi)\,\,\,.
\end{equation}
In Eqs.~(\ref{eq:c1_c2_c3})-(\ref{eq:gamma_prime}) we have used
$C_A\!=\!N_c\!=\!3$, $C_F\!=\!(N_c^2-1)/2N_c\!=\!4/3$,
$\beta_0\!=\!(11C_A - 2n_f)/3$, $n_f\!=\!5$, while $M_\Phi$ is the
mass of the neutral CP-even Higgs boson ($\Phi=h^0,H^0$) and
$\hat{s}$, $\hat{t}$ and $\hat{u}$ are the parton level Mandelstam
variables for the process $b(p_1) g(p_2) \rightarrow b(-p_3)
\Phi(-p_5)$, defined as $\hat{s}\!=\!(p_1+p_2)^2$,
$\hat{t}\!=\!(p_1+p_3)^2$, and $\hat{u}\!=\!(p_2+p_3)^2$. The other
mass scales in this process include $p_3^2=m_b^2$ and
$p_5^2=M_\Phi^2$. We notice that, following
Ref.~\cite{Kidonakis:2004ib}, we have included in $c_1^{bg\rightarrow
  b\Phi}$ only the scale-dependent pieces of the $\delta(\hat{s}_2)$
corrections. In particular, we do not include in $c_1^{bg\rightarrow
  b\Phi}$ the full virtual corrections. We have estimated the impact
of the scale-independent terms using the full one-loop virtual
corrections calculated in Ref.~\cite{Campbell:2002zm} in the
$m_b\!=\!0$ limit, and we have found them to be very small.

In terms of the $c_1$, $c_2$, and $c_3$ defined in
Eq.~(\ref{eq:c1_c2_c3}), the NLO-NLL corrected (total or differential)
cross-section can be written as:
\begin{equation}
\label{eq:nlo-nll}
\hat{\sigma}^{(1)} = \hat{\sigma}^{\text{Born}} \frac{\alpha_s}{\pi}
\{ c_3 \mathcal{D}_1(\hat{s}_2) + c_2 \mathcal{D}_0(\hat{s}_2)
 + c_1 \delta(\hat{s}_2) \}\,\,\,,
\end{equation}
where $\hat{\sigma}^{\text{Born}}$ is the parton level tree level
cross-section (total or differential) for $bg\rightarrow b\Phi$. For
completeness, the tree level differential cross-section for
$bg\rightarrow b\Phi$ can be written as:
\begin{eqnarray} 
\label{eq:born}
\hat{s}^2 \frac{d\hat{\sigma}^{\text{Born}}}{d\hat{t} \, d\hat{u}} &=&
\frac{\alpha_s}{\nc (\n2c -1)}
\left(\frac{\overline{m}_b}{v}\right)^2 \\ &\times & 
\frac{1}{\hat{s}(\hat{t} - m_b^2)^2}
\left\{
- \hat{t} [ M_\Phi^4 + \hat{u}^2 ]
+m_b^2 [ 4 M_\Phi^4 - \hat{s}^2 - 2\hat{u}(M_\Phi^2 + \hat{s} ) ]
-m_b^4 [ 6 \hat{s} + 3 \hat{t} + 4 \hat{u} ]
\right\}\,\,\,,\nonumber
\end{eqnarray}
where the Mandelstam variables in the partonic system were defined
above and $m_b=m_b^\text{pole}$. We notice that we have treated the
final state bottom quark as massive in the $bg\rightarrow b\Phi$ tree
level cross-section. The result in Eq.~(\ref{eq:born}) matches the
well known result~\cite{Campbell:2002zm} in the limit of vanishing
bottom-quark mass. There are many reasons to keep the final state
bottom quark massive in this calculation. The prefactor in
Eq.~(\ref{eq:born}) clearly shows that the bottom quark mass regulates
the small transverse momentum region of the born level differential
cross-section, allowing integration down to zero transverse momentum
-- an important improvement in our treament of this process. The
initial state bottom quark is left massless to be consistent with its
treatment in the evolution of the parton distribution functions. This
is consistent with current treatment of the heavy flavor
thresholds\cite{Aivazis:1993pi, Kramer:2000hn, Chuvakin:2000qc,
  Chuvakin:2001ge, Thorne:2006qt} in our chosen PDF sets and is also
consistent with the implementation of the splitting functions for the
energy evolution which are all strictly massless and assume massless
initial state quarks.

In the same way as the above differential cross-section, the NNLO-NLL
corrected (total or differential) cross-section can be written
as~\cite{Kidonakis:2004ib},
\begin{eqnarray}
\label{eq:nnlo-nll}
\hat{\sigma}^{(2)} &=& \hat{\sigma}^{\text{Born}}
\left( \frac{\alpha_s(\mu_R)}{\pi} \right)^2
\left\{\frac{1}{2} c_3^2 \mathcal{D}_3(\hat{s}_2)
+\left[ \frac{3}{2} c_3 c_2 -\frac{\beta_0}{4} c_3\right] 
 \mathcal{D}_2(\hat{s}_2) \right.\\ \nonumber
&+& \left[ c_3 c_1
  + (C_F+C_A)^2\ln^2\left(\frac{\mu_F^2}{M_\Phi^2}\right)
  -2(C_F+C_A) T_2\ln\left(\frac{\mu_F^2}{M_\Phi^2}\right)
  +\frac{\beta_0}{4} c_3\ln\left(\frac{\mu_R^2}{M_\Phi^2}\right)
  -\zeta_2 c_3^2\right] \mathcal{D}_1(\hat{s}_2) \\ \nonumber
&+&\left[-(C_F+C_A) \ln\left(\frac{\mu_F^2}{M_\Phi^2}\right) c_1
-\frac{\beta_0}{4} (C_F+C_A) \ln\left(\frac{\mu_F^2}{M_\Phi^2}\right)
\ln\left(\frac{\mu_R^2}{M_\Phi^2}\right) 
+(C_F+C_A)\frac{\beta_0}{8} \ln^2\left(\frac{\mu_F^2}{M_\Phi^2}\right)
\right.\nonumber\\
&&\left.\left.\,\,-\,\zeta_2 c_2 c_3 +\zeta_3 c_3^2
\phantom{\frac{1}{2}}\right] \mathcal{D}_0(\hat{s}_2)
\right\}\,\,\,,\nonumber
\end{eqnarray}
while the NNNLO-NLL corrected (total or differential) cross-section can
be written as~\cite{Kidonakis:2005kz},
\begin{eqnarray}
\label{eq:nnnlo-nll}
\hat{\sigma}^3&=&\hat{\sigma}^{\text{Born}}
\left(\frac{\alpha_s(\mu_R)}{\pi}\right)^3
\left\{
\frac{1}{8}c_3^3\mathcal{D}_5(\hat{s}_2)+
\left[\frac{5}{8}c_3^2c_2-\frac{5}{2}c_3X_3\right]\mathcal{D}_4(\hat{s}_2)
\right.\\
&+&\left[
c_3(c_2^\mu)^2+2c_3T_2c_2^\mu+\frac{1}{2}c_3^2c_1^\mu-\zeta_2c_3^3-
4c_2^\mu X_3+2c_3X_2^\mu\right]\mathcal{D}_3(\hat{s}_2)\nonumber\\
&+&\left[
\frac{3}{2}c_3c_2^\mu c_1^\mu
+\frac{1}{2}(c_2^\mu)^3+\frac{3}{2}T_2(c_2^\mu)^2-3\zeta_2c_3^2c_2
+\frac{5}{2}\zeta_3c_3^3
+\frac{27}{2}\zeta_2c_3X_3+3c_2^\mu X_2^\mu
-\frac{3}{2}c_3\left(X_1^{\mu^2}+X_1^\zeta\right)\right]\mathcal{D}_2(\hat{s}_2)
\nonumber\\
&+&\left.\left[
(c_2^\mu)^2c_1^\mu-\zeta_2c_3^2c_1^\mu -\frac{5}{2}\zeta_2c_3
\left((c_2^\mu)^2+2T_2c_2^\mu\right)+5\zeta_3c_3^2c_2^\mu
+12\zeta_2c_2^\mu X_3-5\zeta_2c_3X_2^\mu 
-2c_2^\mu\left(X_1^{\mu^2}+X_1^\zeta\right)\right]\mathcal{D}_1(\hat{s}_2)\right\}
\,\,\,,\nonumber
\end{eqnarray}
where $\zeta_2\!=\!\zeta(2)\!=\!\pi^2/6$,
$\zeta_3\!=\!\zeta(3)\!=\!1.2020569...$, $c_1^\mu$ and $c_2^\mu$ are
the scale dependent parts of $c_1$ and $c_2$ defined in
Eq.~(\ref{eq:c1_c2_c3}), $T_2$ is defined in Eq.~(\ref{eq:t2}), while
$X_i$, $X_i^\mu$, $X_i^{\mu^2}$, and $X_i^\zeta$ (for $i=1,2,3$) are
functions of the kinematic variables of the process described in
Refs.~\cite{Kidonakis:2004ib} and \cite{Kidonakis:2005kz}.

It is important to understand the limitations of this
calculation. Indeed, the kinematics of the resummed total or
differential cross-section is fixed by the tree-level process (see
Eqs.~(\ref{eq:nlo-nll})-(\ref{eq:nnnlo-nll})) and it is therefore a
$2\rightarrow 2$ kinematic, even upon inclusion of resummed
higher-order QCD corrections. At the same time, the observation of
Higgs boson production with one $b$-jet is subject to identification
cuts imposed on the final state $b$-jet transverse momentum ($p_T^b$)
and pseudo-rapidity ($\eta_b$), which translate, given the
$2\rightarrow 2$ kinematic of the resummed approach, into cuts on
$p_T^\Phi$, at all orders of the resummed QCD corrections.  This has
to be taken into account when looking at distributions, in particular
the $p_T^\Phi$ distribution. Therefore, we interpret the corresponding
$p_T^\Phi$ distribution as an improved estimate of the corresponding
fixed-order result in the region of low $p_T^\Phi$ above the $p_T^b$
cut. If the experimental cut on $p_T^b$ is lowered, larger and larger
portions of the resummed distributions become important, since one
enters more and more the region where the resummed corrections are
important and a fixed-order calculation is not expected to give a
reliable result. In spite of the fact that the resummed $p_T^\Phi$
distribution does not include effects from gluon dynamics, the
bottom-quark mass does much to stabilize the small transverse momentum
region and gluon dynamics effects should be under control in this
region.

There is an additional complication when we compare our resummed
results to the fixed-order results produced by MCFM. The way in which
MCFM calculates the NLO corrections to Higgs plus bottom quark
processes is broken into three parts, as described in the MCFM
manual\cite{mcfm}. The process labeled $143$ takes into account a
Higgs produced with two additional bottom quarks, both of which are
observed. This process would need to be treated separately in our
resummation formalism. The unaccounted for process has physical
significance for our signal if at least one of the bottom quarks were
to meet the identification cuts. Therefore we have added this process
to our resummed results. This addition makes our hybrid curves highly
reliable in all regions.

\section{Results}
\label{sec:results}

In this section we present several results that will illustrate both
formal and phenomenological aspects of the resummed calculation.  We
always show results for $bh^0$ production at the Tevatron and for
$bH^0$ production at the LHC. The final state bottom quark is
identified imposing that its transverse momentum and pseudorapidity
satisfy: $p_T^b > p_T^{b,\textrm{cut}}$ (where $p_T^{b,\textrm{cut}}$
will be specified separately for different plots) while $|\eta^b|<2$
at the Tevatron and $|\eta^b|<2.5$ at the LHC. We will specify when no
identification cuts are imposed on the resummed calculation and
further justify our choice.

\begin{figure}
\begin{center}
  \includegraphics[bb=50 50 410 300,scale=0.95]{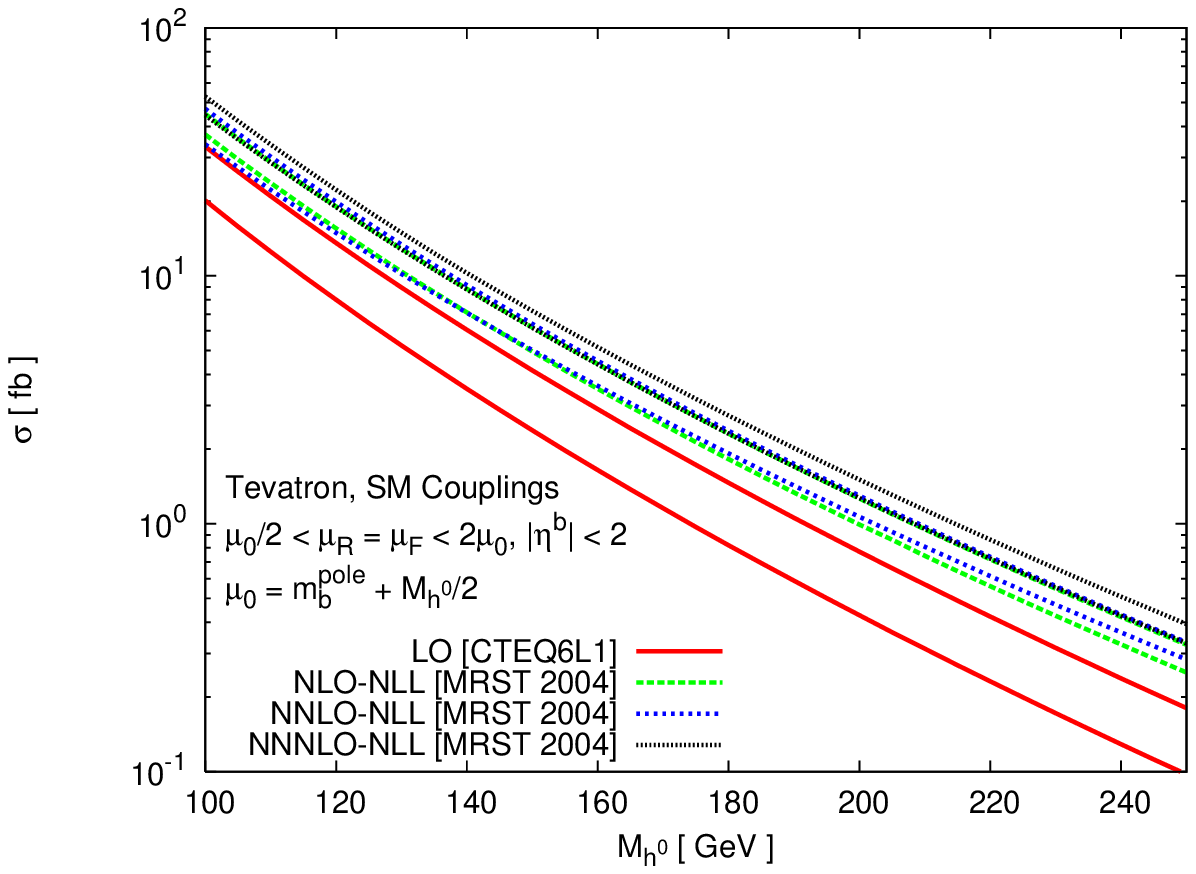} 
  \includegraphics[bb=470 -250 770 350,scale=0.4]{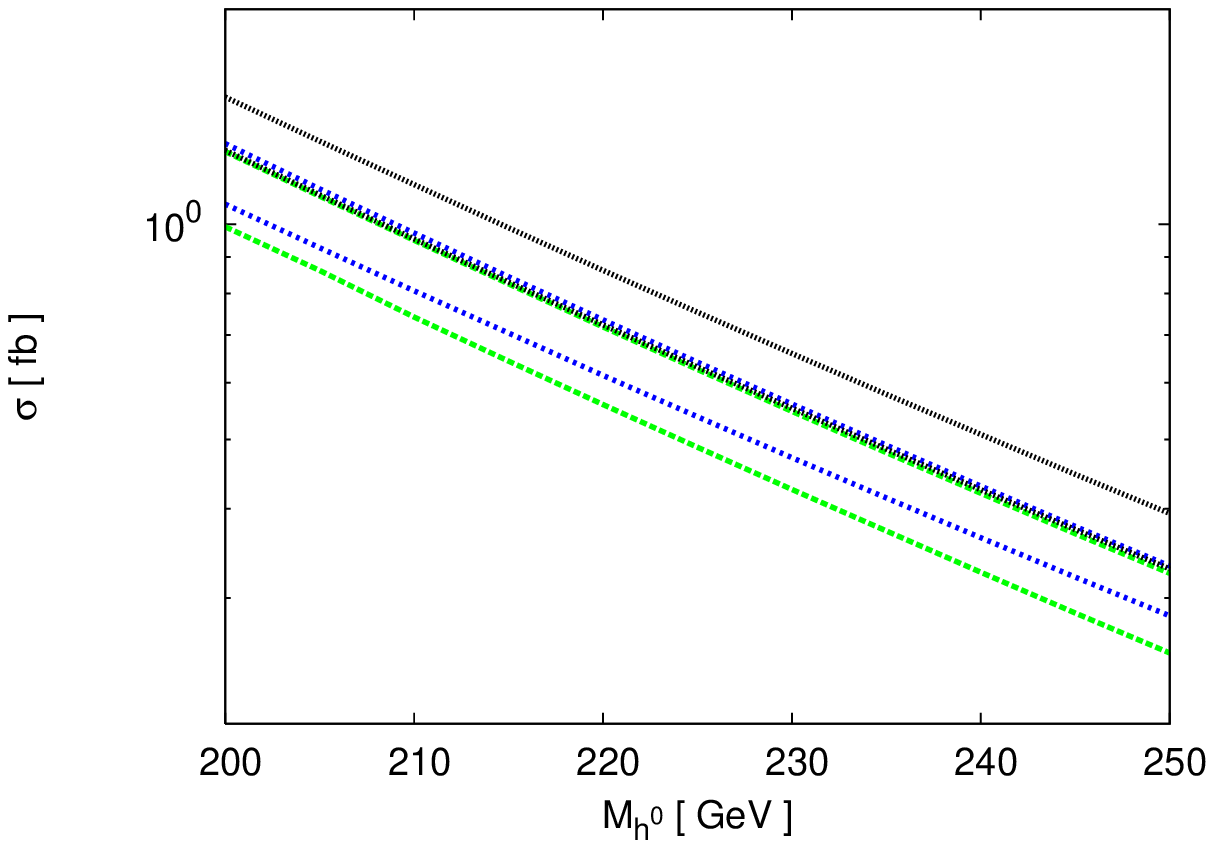} \\
  \includegraphics[bb=50 50 410 300,scale=0.95]{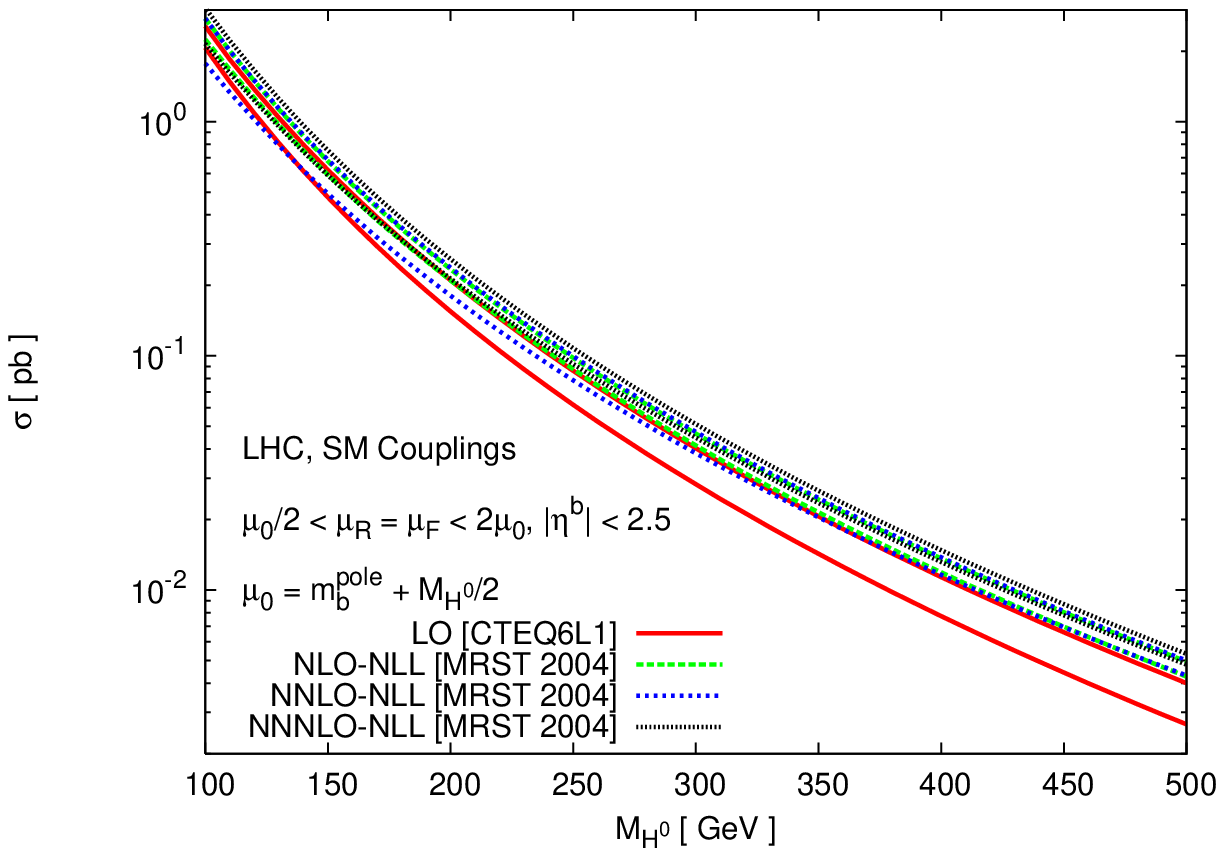}
  \includegraphics[bb=470 -250 770 350,scale=0.4]{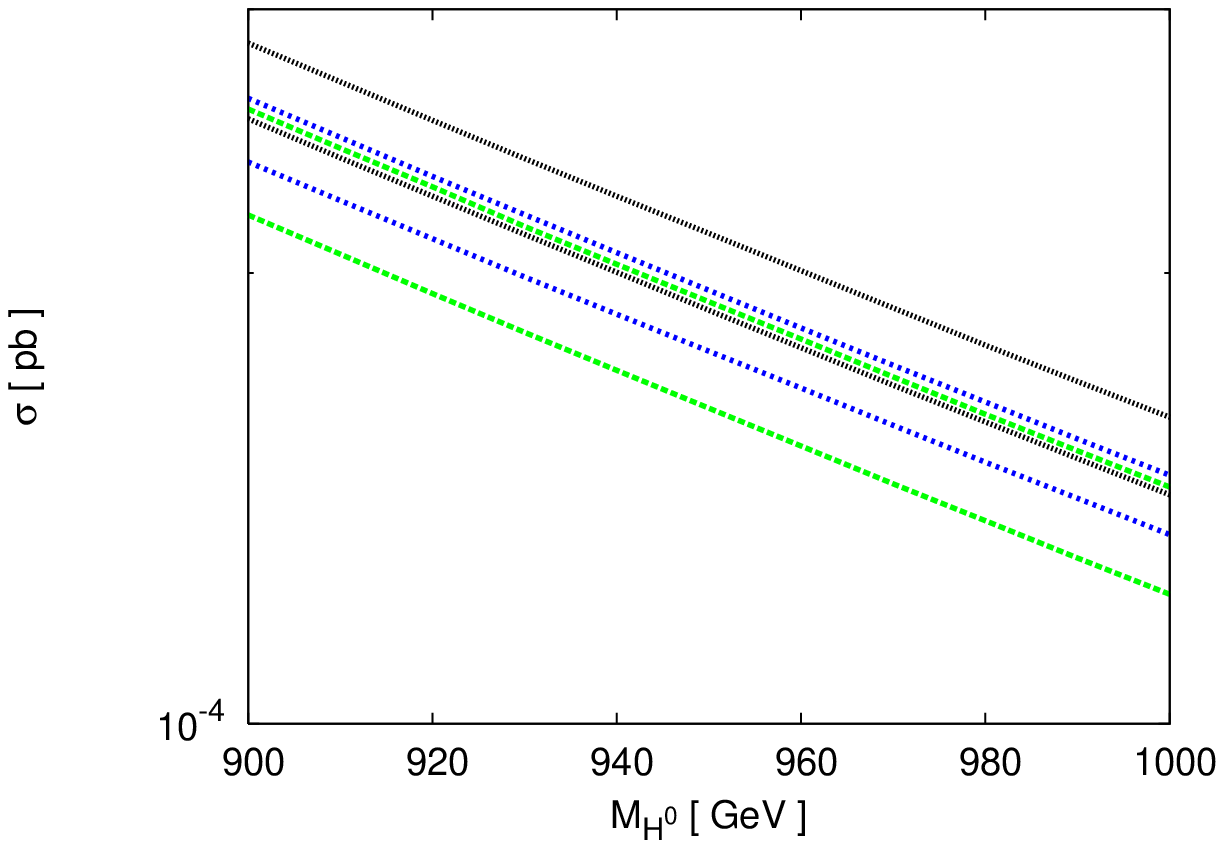} 
\end{center}
\caption[]{A comparison of different perturbative orders of the
  resummed total cross-sections versus Higgs mass when the
  factorization $\mu_F$ and renormalization $\mu_R$ scales are varied
  by a factor of two about the central value $\mu_F=\mu_R=\mu_0$ with
  $\mu_0=m_b^{\text{pole}}+M_\Phi/2$ ($\Phi=h^0,H^0$). Each band shows
  the maximum variation for a given perturbative order. The inlays
  illustrate in greater detail the large $M_\Phi$ region. We have
  added the additional final state $b\bar{b}\Phi \, (\Phi = h^0, H^0)$
  to our resummed results as described in the text.}
\label{fig:mu_bands} 
\end{figure}

With the exception of Fig.~\ref{fig:mu_bands}, where we study the
renormalization and factorization scale dependence of the resummed
cross-section, we set the renormalization and factorization scales to
$\mu_R\!=\!\mu_F\!=\!\mu_0/2$, where
$\mu_0=(m_b^\text{pole}+M_{\Phi}/2)$\footnote{Several studies indicate
  that $\mu_F\simeq M_\Phi/4$ is the most appropriate choice of
  factorization scale for the process studied in this
  paper~\cite{Rainwater:2002hm,Plehn:2002vy,Boos:2003yi,Maltoni:2003pn}.}. LO
results are obtained using CTEQ6L1 PDFs~\cite{Pumplin:2002vw}, 1-loop
$\alpha_s(\mu_R)$ and 1-loop bottom-quark running mass
($\overline{m}_b(\mu_R)$) (see Eq.~(\ref{eq:mb_ms_1l_2l})). NLO
results are obtained using the NLO set of MRST 2004
PDFs~\cite{Martin:2004ir} (since they also provide NNLO PDFs), 2-loop
$\alpha_s(\mu_R)$ and 2-loop $\overline{m}_b(\mu_R)$ (see
Eq.~(\ref{eq:mb_ms_1l_2l})). Finally NNLO and NNNLO results are
obtained using the NNLO set of MRST PDFs~\cite{Martin:2004ir}, as well
as 3-loop and 4-loop $\alpha_s(\mu_R)$ and
$\overline{m}_b(\mu_R)$~\cite{Vermaseren:1997fq}
respectively. Although the MRST group has produced a LO PDF
set~\cite{Martin:2002dr}, it is from an older fit to data than the
CTEQ6L1 PDF set. We also note that the available LO MRST 2001 set
includes NLO and NNLO PDFs which differ from the MRST 2004 NLO and
NNLO sets, as expected, and keeping a consistent PDF set is no longer
viable option if modern PDFs are required. Moreover, as our lowest
order results are meant for strictly illustrative purposes, the choice
of modernity over uniformity is a minor one.

\begin{figure}
\begin{center}
  \includegraphics[scale=0.95]{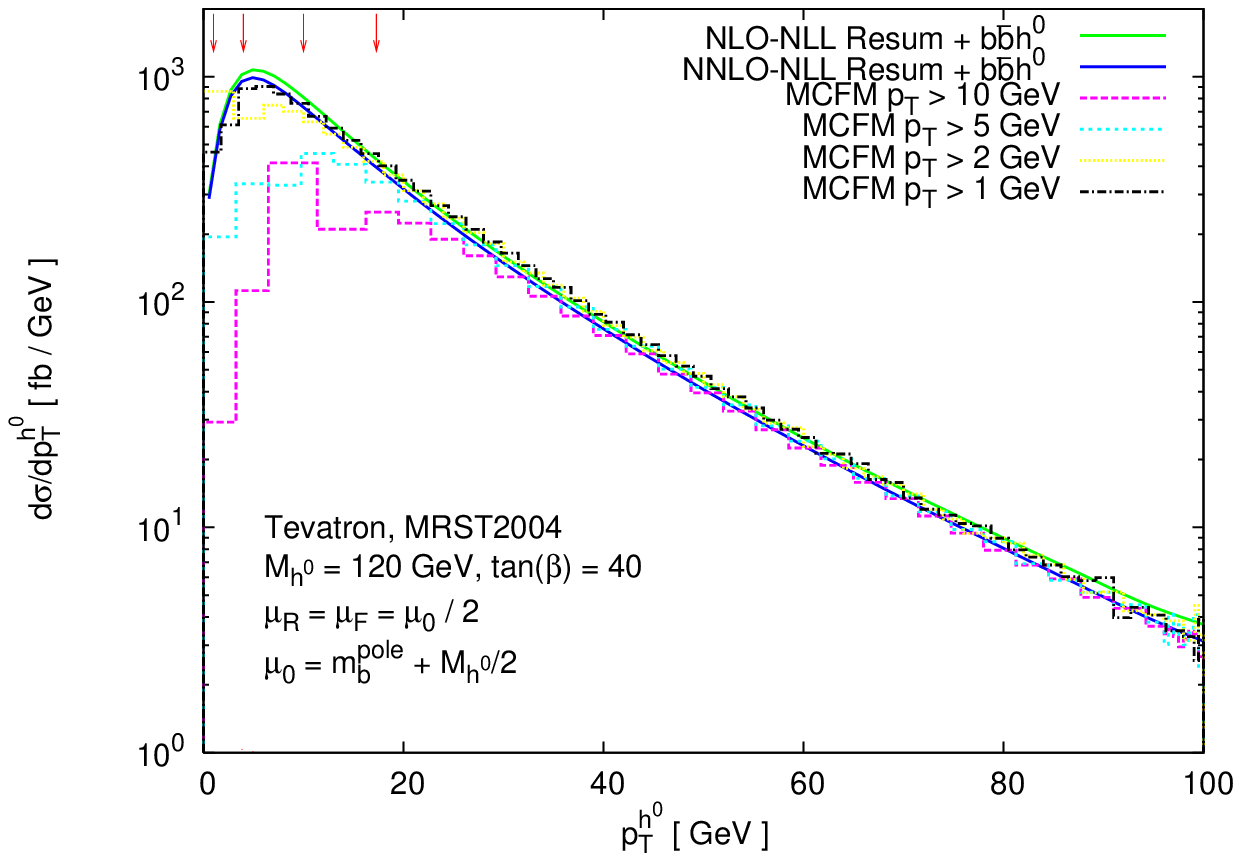} \\
  \includegraphics[scale=0.95]{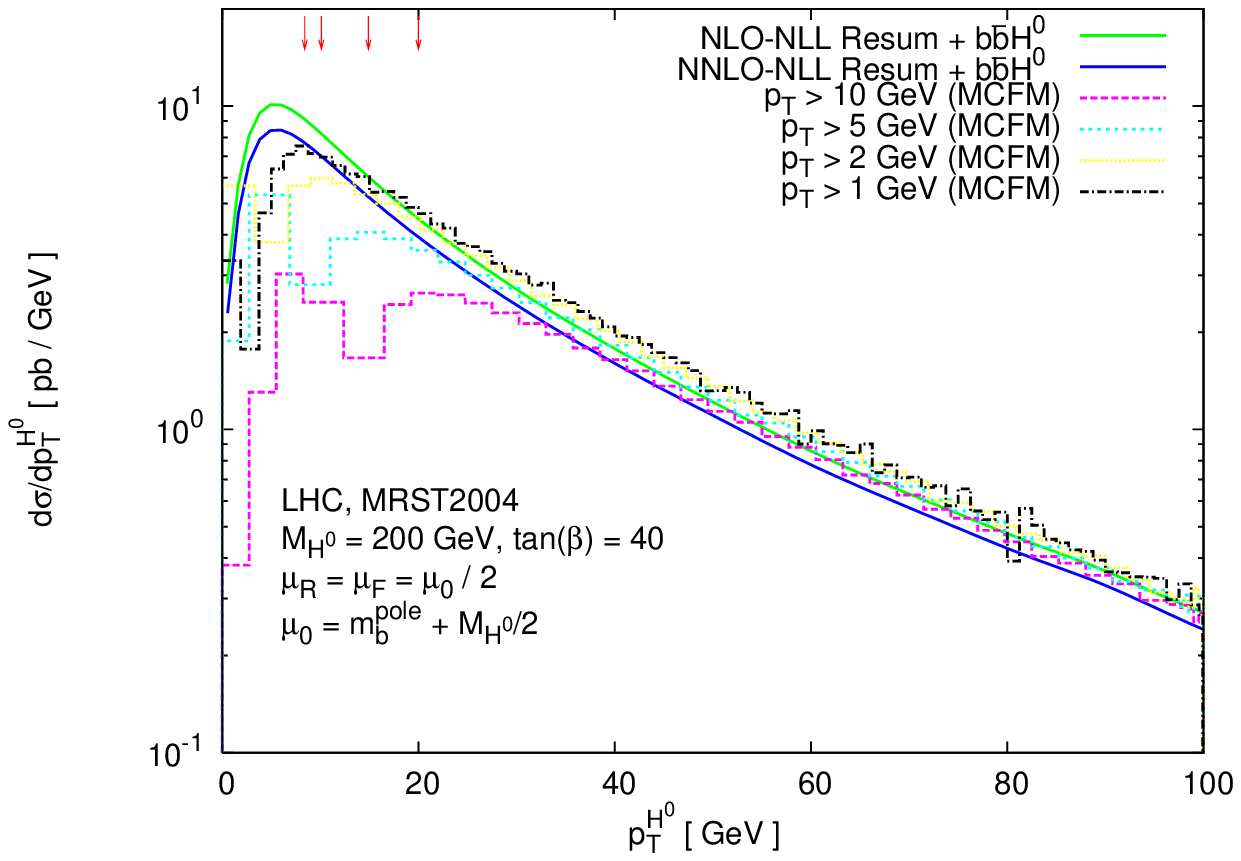}
\end{center}
\caption[]{Comparison of the NLO fixed-order 5FNS and the NLO-NLL and
  NNLO-NLL resummed transverse momentum distribution at the Tevatron
  and the LHC for different cuts in the transverse momentum of the
  final state bottom quark in the fixed-order calculation. We have
  added the additional final state $b\bar{b}\Phi \, (\Phi = h^0, H^0)$
  to our resummed results as described in the text. The arrows at the
  top of the graph guide the eye to show where each of the fixed-order
  distributions peak when the histogram is smoothed. As expected,
  lower cuts on the transverse momentum lead to peaks at smaller
  values. The fixed-order differential cross-section also begins to
  pick up more of the large soft contributions from the region of
  small transverse momentum. Here we have set $M_{h^{0}} = 120$~GeV
  (Tevatron), $M_{H^{0}} = 200$~GeV (LHC), and $\mu_R=\mu_F=\mu_0/2$
  with $\mu_0=(m_b^{\text{pole}} + M_{h^0/H^{0}}/2)$.}
\label{fig:dsigma_dptphi}
\end{figure}

\begin{figure}
\begin{center}
  \includegraphics[scale=0.95]{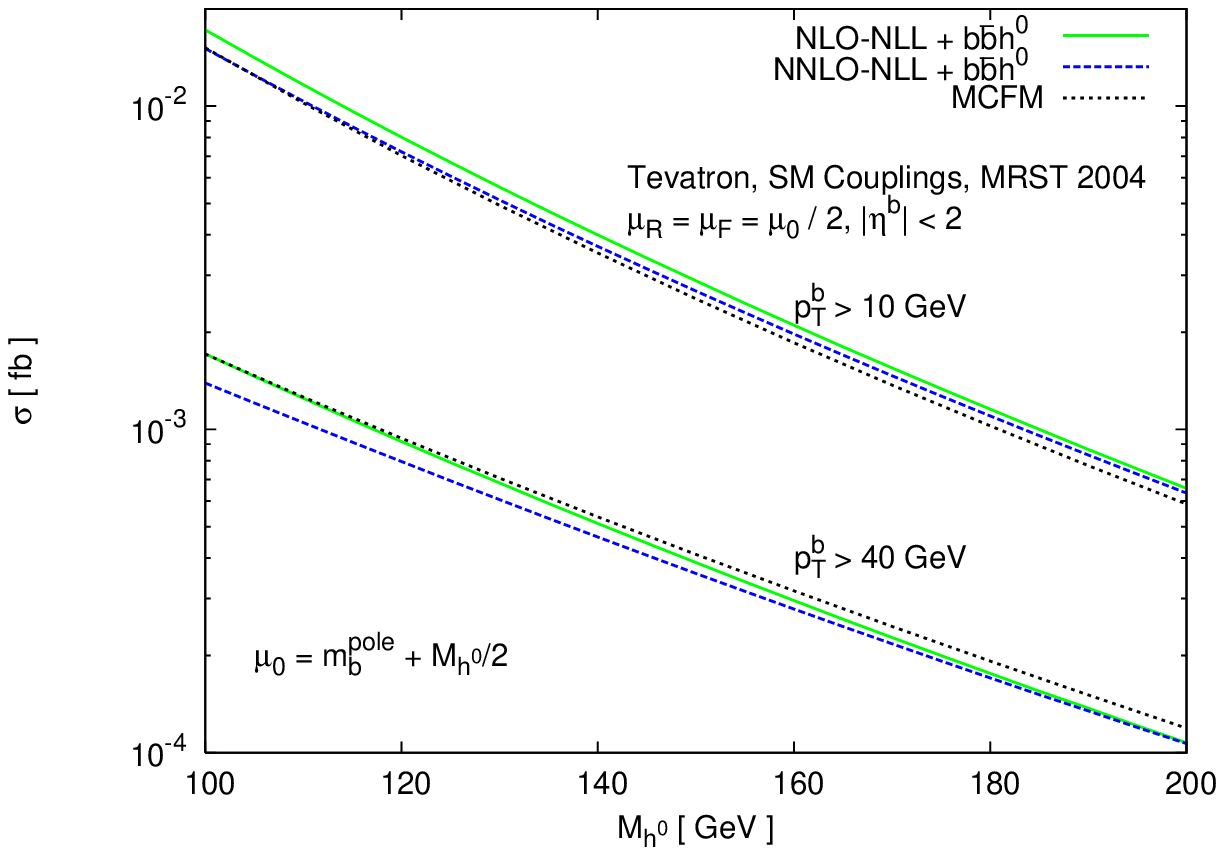} \\
  \includegraphics[scale=0.95]{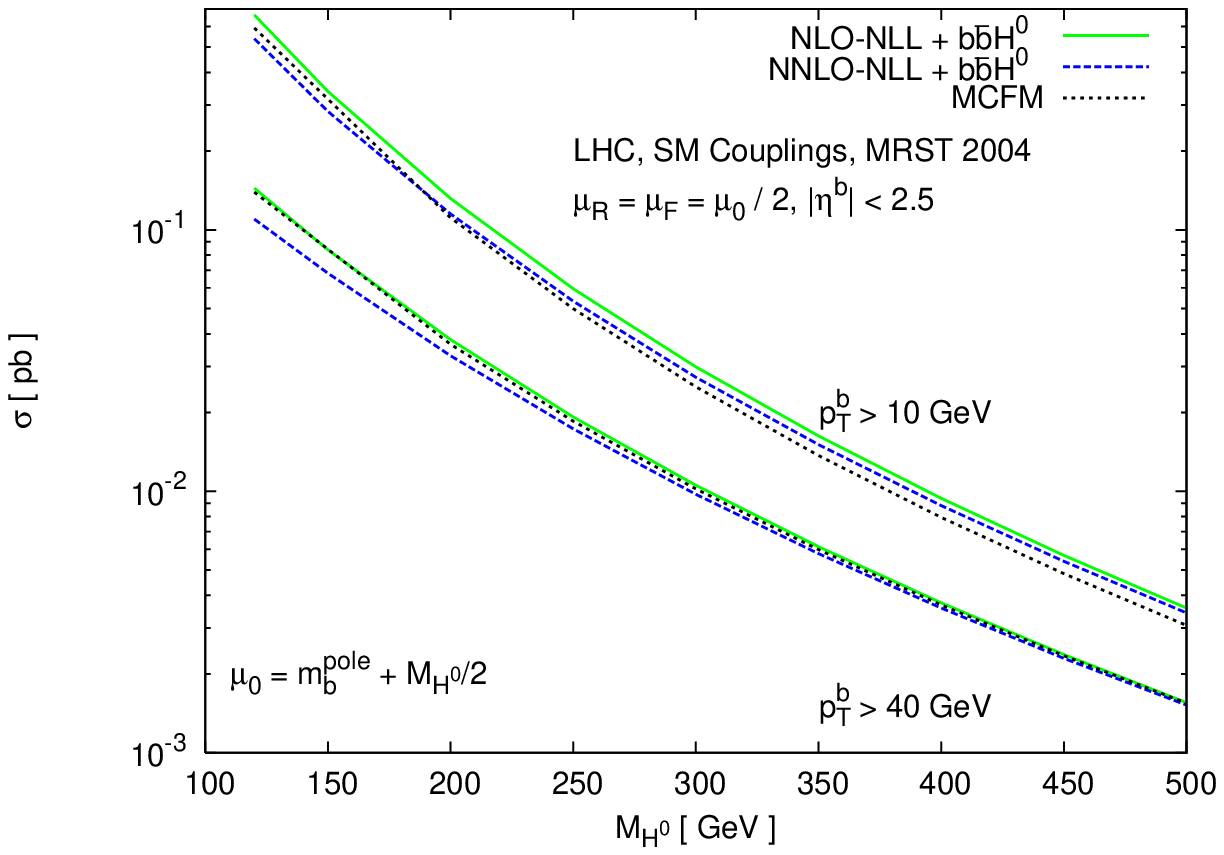}
\end{center}
\caption[]{A comparison of the resummed NLO-NLL and NNLO-NLL total
  cross-sections with the fixed-order NLO corrections as calculated by
  MCFM (5FNS), for $p_{T}^b > 10$~GeV and $p_{T}^b > 40$~GeV, with
  varied Higgs boson masses. We have added the additional final state
  $b\bar{b}\Phi \, (\Phi = h^0, H^0)$ to our resummed results as
  described in the text.}
\label{fig:sigma_vs_mh_ptcut_10_40}
\end{figure}

\begin{figure}
\begin{center}
  \includegraphics[scale=0.95]{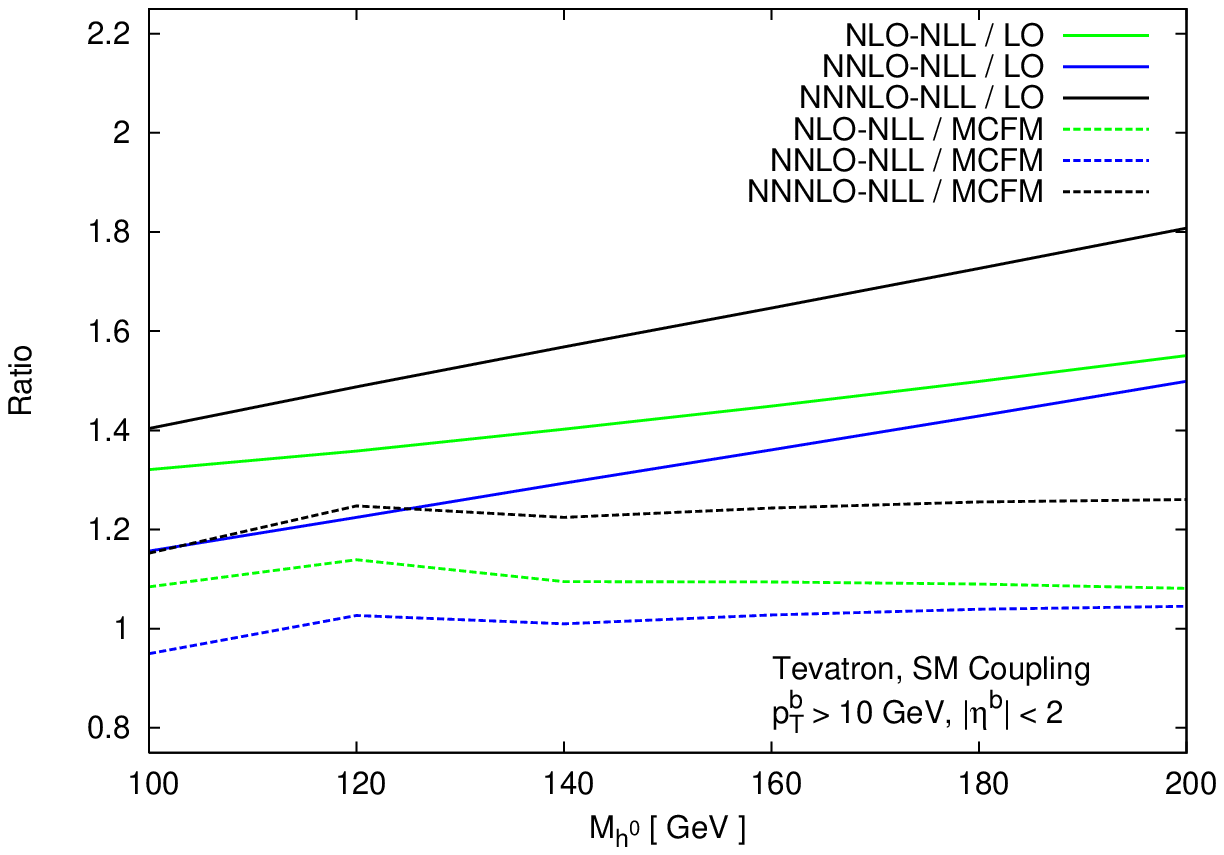} \\
  \includegraphics[scale=0.95]{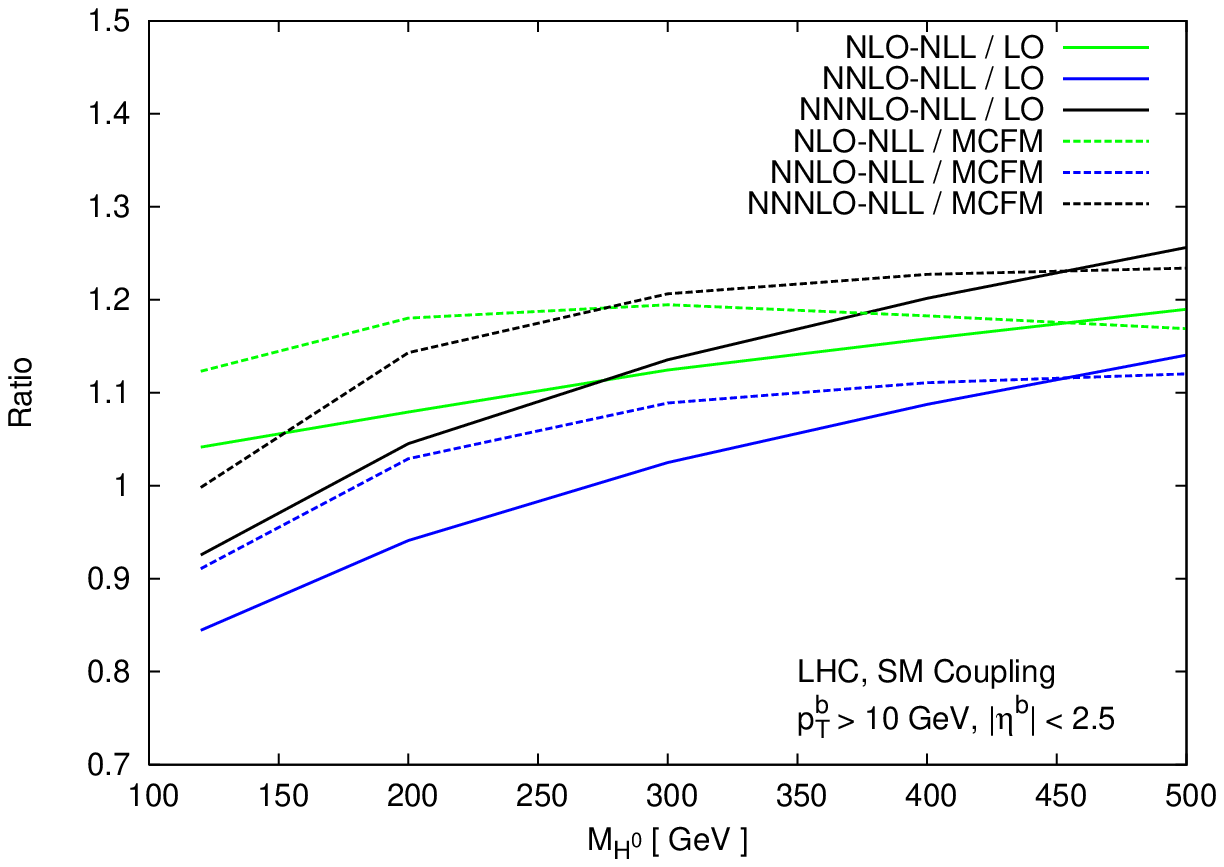}
\end{center}
\caption[]{Ratios of the NLO-NLL, NNLO-NLL, and NNNLO-NLL to the LO
  total cross-section for different values of the Higgs boson mass
  ($M_{h^0}$ at the Tevatron and $M_{H^0}$ at the LHC), and
  $\mu_R=\mu_F=\mu_0/2= (m_b^{\text{pole}} + M_{h^0,H^0}/2)/2$. We
  have added the additional final state $b\bar{b}\Phi \, (\Phi = h^0,
  H^0)$ to our resummed results as described in the text.}
\label{fig:kfactors}
\end{figure}

\begin{figure}
\includegraphics[scale=1.00]{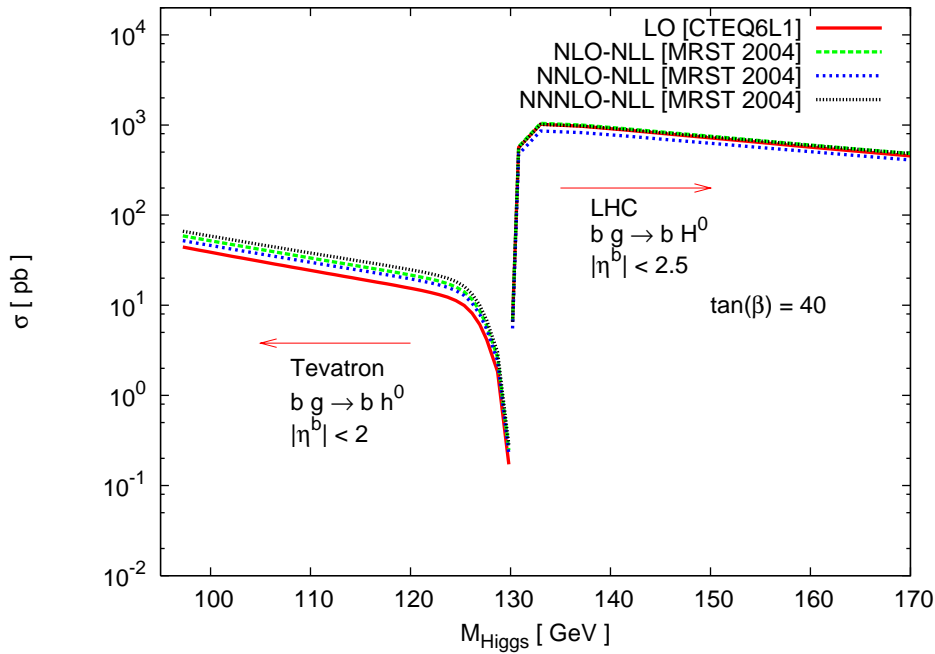} \\
\caption[]{Total MSSM cross-sections for the two processes, $bg
  \rightarrow g h^{0}$ and $bg \rightarrow bH^{0}$, at the Tevatron
  and the LHC respectively, as a function of $M_{h^0,H^0}$, when
  $\mu_R=\mu_F=\mu_0/2= (m_b^{\text{pole}} + M_{h^0,H^0}/2)/2$.  We
  have added the additional final state $b\bar{b}\Phi \, (\Phi = h^0,
  H^0)$ to our resummed results as described in the text.}
\label{fig:mssm} 
\end{figure}

Figs.~\ref{fig:dsigma_dptphi} and \ref{fig:mssm} are obtained using
MSSM couplings, since they are of direct phenomenological interest.
In particular we use the setup explained at the end of
Section~\ref{sec:mssm}, with $\tan\beta=40$. Different $M_{h^0,H^0}$
hence correspond to different values of $M_{A^0}$. On the other hand,
the perturbative properties of the resummed cross-sections (see
Figs.~\ref{fig:sigma_vs_mh_ptcut_10_40}, \ref{fig:mu_bands} and
\ref{fig:kfactors}) are studied using SM couplings.

In Fig.~\ref{fig:dsigma_dptphi} we plot the $p_T^\Phi$
($\Phi=h^0,H^0$) distribution for Higgs production with one bottom
quark at both the Tevatron and the LHC, for $M_{h^0}=120$~GeV and
$M_{H^0}=200$~GeV respectively.  We use MRST 2004 PDFs for both
fixed-order and resummed results.  The fixed-order results are
obtained using MCFM~\footnote{The MCFM curves are a combination of the
processes labelled in MCFM as $141$ ($\Phi bg$, $\Phi=h^0,H^0$), $142$
($\Phi b(\bar{b})$), and $143$ ($\Phi b\bar{b}$). For more details see
\cite{mcfm}.}.  The NLO-NLL and NNLO-NLL resummation results, obtained
with no $p_T^b$ identification cut imposed on the final state bottom
quark, are compared to the NLO 5FNS fixed-order results with
decreasing values of the $p_T^b$ identification cut ($p_T^b > 10,5,2$,
and $1$~GeV). As we already commented at the end of
Section~\ref{sec:resummation}, the resummed cross-section in
Eqs.~(\ref{eq:nlo-nll})-(\ref{eq:nnnlo-nll}) is bound to a
$2\rightarrow 2$ kinematic. A cut on $p_T^b$ automatically translates
into a cut in $p_T^\Phi$ and truncates the $p_T^\Phi$ spectrum below
that point. Therefore, in order to obtain the $p_T^\Phi$ distribution
over the entire $p_T^\Phi$ range, we do not impose a cut on the
$p_T^b$ of the resummed differential cross-section with the caveat
that fixed-order distributions obtained for a given cut on $p_T^b$
have to be compared with the resummed distribution for
$p_T^\Phi>p_T^{b,\textrm{cut}}$.

Overall, even for sizable cuts on $p_T^b$ (say $p_T^b>10$~GeV), we can
see very good agreement between the NLO fixed-order calculation and
the NLO-NLL resummed differential cross-section well below the mass
scale of the process ($M_{h^{0}}$ or $M_{H^{0}}$), and we begin to see
an appreciable difference starting at $\mu_R=\mu_F = \mu_0/2 =
(m_b^{\text{pole}} + M_{\Phi}/2 )/2$, as expected. The resummed
distribution seems to interpolate between the fixed-order curves,
smoothing the low $p_T^\Phi$ portion of the spectrum where the large
soft contributions start to be more relevant. Higher-order terms in
the perturbative expansion of the resummed cross-section should then
represent better approximations of the $p_T^\Phi$ spectrum in the low
momentum region, where most of the statistics are accumulated.  We
then interpret the NLO-NLL and the NNLO-NLL curves as improvements
over the fixed-order calculation in the region of low $p_T^\Phi$ above
the experimental $p_T^b$ cut.  For the choice of parameters (Higgs
boson masses and renormalization/factorization scale) and PDFs in
Fig.~\ref{fig:dsigma_dptphi} the NLO-NLL and NNLO-NLL curves are
remarkably close. For other choices of parameters they typically have
the same shape, but they can have more distinguishable magnitudes.
This is better illustrated in Fig.~\ref{fig:mu_bands} where we study
the dependence on the renormalization and factorization scales of the
various terms in the expansion of the resummed cross-section.

We have also compared the total cross-sections for $bh^0$ and $bH^0$
production when a $p_T^b$ identification cut is imposed on both
fixed-order and resummed calculations. In the resummed case, this
corresponds to dropping the region of the $p_T^\Phi$ spectrum below
$p_T^b$. By looking at Fig.~\ref{fig:dsigma_dptphi} we do not expect
the fixed-order and resummed results to agree well when large $p_T^b$
cuts are imposed. Results are shown in
Fig.~\ref{fig:sigma_vs_mh_ptcut_10_40}, where the total cross-section
is plotted as a function of $M_\Phi$ ($\Phi=h^0,H^0$), for
$p_T^b>10$~GeV and $p_T^b>40$~GeV respectively, at both the Tevatron
and the LHC. The Tevatron plots, in particular, confirm our
expectations. We also notice that the NNLO-NLL and NLO-NLL results
converge better the larger the Higgs boson mass, i.e. when the
kinematic approaches the threshold condition. This last aspect of the
perturbative behavior of the resummed cross-section is strongly
confirmed in Fig.~\ref{fig:mu_bands}.

Indeed, Fig.~\ref{fig:mu_bands} shows the perturbative behavior of the
resummed cross-section expanded at LO, NLO-NLL, NNLO-NLL and
NNNLO-NLL, when the renormalization and factorization scales are
varied between $\mu_0/2$ and $2\mu_0$, for $\mu_0=m_b^{\text{pole}} +
M_\Phi/2$ ($\Phi=h^0,H^0$). We keep $\mu_F=\mu_R$ in our variation
since this is how PDF packages are structured and varying the two
scales separately would cause a mismatch at this level. The LO,
NLO-NLL and NNLO-NLL bands are completely consistent, since all
perturbative quantities ($\alpha_s(\mu_R)$, $\overline{m}_b(\mu_R)$
and PDFs) are defined at the correct perturbative order. The NNNLO-NLL
cannot be matched to the correct perturbative order PDFs, since they
are not available at NNNLO. No $p_T^b$ identification cut has been
used in this plot, since we are only interested in the theoretical
behavior of the cross-section as a function of the scale. For the same
reason, we have used SM couplings. We notice a large impact in going
from LO to the first order of QCD corrections, i.e. NLO-NLL, while the
NNLO-NLL order add only a small variation. The large difference in
going from LO to NLO could in part be due to the different set of PDFs
used.  On the other hand, within the theoretical uncertainty due to
the residual scale dependence, the NLO-NLL and NNLO-NLL predictions
are completely consistent. The residual uncertainty in the NNLO-NLL
prediction is smaller than the corresponding uncertainty in the
NLO-NLL prediction for large Higgs boson masses (see inlays in
Fig.~(\ref{fig:mu_bands})), i.e. in the threshold region, where the
resummed formalism works better.

All of these characteristics neatly illustrate
the correct behavior of the resummed perturbative series.  As far as
the NNNLO-NLL corrections go, they do not seem to follow this pattern
of compatibility with the lower order result and
reduction of the residual scale dependence, but we need to
remember that there is a mismatch with the perturbative order of the
PDFs. Because of this, we would suggest to limit the improvement of the
NLO fixed-order cross-section to the NNLO-NLL resummed predictions.
This is indeed what we show in the distribution plots of
Fig.~\ref{fig:dsigma_dptphi}.

In Fig.~\ref{fig:kfactors}, we summarize our results by plotting
various \emph{K-factors}, i.e. ratios of the higher-order (NLO-NLL,
NNLO-NLL, and NNNLO-NLL) to the LO cross-section, as a function of the
Higgs boson mass.  We also give ratios of the various orders of the
resummed cross-section to the NLO fixed-order cross-section,
calculated with MCFM. All higher-order results use MRST 2004 PDFs,
while the LO ones use CTEQ6L1.  All curves on Fig.~\ref{fig:kfactors}
are obtained for $\mu_R=\mu_F=\mu_0/2$ for $\mu_0=m_b^{\text{pole}} +
M_\Phi/2$.  These curves quantify the effect of the higher-order
corrections to the total cross-section, as well as to the fixed-order
cross-section.  We see that the threshold effects are more relevant at
the Tevatron than the LHC for these particular parameter choices, in
that the ratios deviates more from unity at the Tevatron than at the
LHC.

Finally, in Fig.~\ref{fig:mssm} we emphasize the behavior of the $bh^0$
and $bH^0$ cross-sections in the MSSM, by plotting both production
cross-sections in the same plot as a function of $M_\Phi$
($\Phi=h^0,H^0$). The specific behavior of the MSSM Higgs boson masses
and couplings is very well represented.  As the lightest Higgs scalar
boson ($h^0$) becomes inaccessible, the heavier scalar Higgs boson
($H^0$) turns on and begins being produced. The effects are on top of
the QCD corrections, that are here represented in terms of the various
terms in the perturbative expansion of the resummed cross-section.
The results plotted in this figure have been obtained without imposing
a specific identification cut on the final state bottom quark, since the
plot only aims at showing at a glance the effect of adding leading
higher-order terms in the calculation of the total cross-section. It
is clear that all QCD effects should be included in any realistic
attempt to find a Higgs boson(s) in nature in the coming years.

\section{Conclusions}
\label{sec:conclusions}

We apply the threshold resummation formalism to improve the NLO
fixed-order predictions for the production cross-sections of a
MSSM Higgs boson ($h^0,H^0$) in association with one bottom quark.
We focus in particular on the Higgs boson transverse momentum
distribution and study how to use the the threshold resummation
formalism to provide a more accurate prediction in the region of low
transverse momentum. We also study in detail the perturbative behavior
of the resummed cross-section and establish its limits and validity.

We see that the threshold corrections are important at both the
Tevatron and the LHC, in particular for large Higgs boson masses,
because the Higgs would be produced closer to threshold where the
resummation effects are greatest. As expected, at both colliders the
most relevant impact is in the low portion of the $p_T^\Phi$
distribution, where also most of the statistics are
accumulated. Therefore, we expect these results to provide valuable
information to experiments searching for evidence of MSSM scalar Higgs
bosons at both the Tevatron and the LHC.

\begin{acknowledgments}

The authors would like to thank Nikolaos Kidonakis for clarifying
several issues in the application of the threshold resummation
formalism.  This work is supported in part by the U.S. Department of
Energy under grants DE-FG02-97IR41022 (L.R. and B.J.F.) and
DE-AC02-98CH10886 (C.B.J.). One author (B.J.F.) would like to thank the
hospitality of the Fermilab theory group where much of this work
occurred.

\end{acknowledgments}
\bibliography{resum-bjet}
\end{document}